\RequirePackage[2020-02-02]{latexrelease}
\documentclass[runningheads]{llncs}
\usepackage[T1]{fontenc}
\usepackage{graphicx}
\usepackage{multirow}
\usepackage{lscape}
\usepackage{graphicx}
\usepackage{caption}
\usepackage{subcaption}
\usepackage{enumerate}
\usepackage{enumitem}
\usepackage{microtype}

\usepackage{xcolor}
\usepackage[printwatermark]{xwatermark}
\usepackage{tikz}
\newsavebox\mybox
\savebox\mybox{\tikz[color=black,opacity=1]\node[text width=15cm,align=center]{Accepted at GameSec 2022 - Conference on Decision and Game Theory for Security.};}
\newwatermark*[
 page=1,
 angle=0,
 scale=.75,
 xpos=0,
 ypos=130
]{\usebox\mybox}

\begin{document}
\title{Using Deception in Markov Game to Understand Adversarial Behaviors through a Capture-The-Flag Environment}
\titlerunning{Using Deception to Understand Adversarial Behaviors}

\author{Siddhant Bhambri\inst{1\dag} \and
Purv Chauhan\inst{1\dag} \and
Frederico Araujo\inst{2} \and
Adam Doup\'e\inst{1} \and
Subbarao Kambhampati\inst{1}}
\authorrunning{S. Bhambri \& P. Chauhan et al.}
\institute{Arizona State University, Tempe, AZ, USA\\
\email{\{sbhambr1, prchauha, doupe, rao\}@asu.edu}\\
\and
IBM Research, Yorktown Heights, NY, USA\\
\email{frederico.araujo@ibm.com}}
\maketitle              % typeset the header of the contribution
\def\thefootnote{\dag}\footnotetext{These authors contributed equally to this work.}
\begin{abstract}
Identifying the actual adversarial threat against a system vulnerability has been a long-standing challenge for cybersecurity research. To determine an optimal strategy for the defender, game-theoretic based decision models have been widely used to simulate the real-world attacker-defender scenarios while taking the defender's constraints into consideration. In this work, we focus on understanding human attacker behaviors in order to optimize the defender's strategy. To achieve this goal, we model attacker-defender engagements as Markov Games and search for their Bayesian Stackelberg Equilibrium. We validate our modeling approach and report our empirical findings using a Capture-The-Flag (CTF) setup, and we conduct user studies on adversaries with varying skill-levels. Our studies show that application-level deceptions are an optimal mitigation strategy against targeted attacks---outperforming classic cyber-defensive maneuvers, such as patching or blocking network requests. We use this result to further hypothesize over the attacker's behaviors when trapped in an embedded honeypot environment and present a detailed analysis of the same.

\end{abstract}
\section{Introduction}

Cybersecurity research, particularly that focused on finding optimal decision strategies for a system defender, when faced by an adversarial threat, has almost always involved a strong dependence on the assumptions made over the adversary's capabilities and the associated threat posed on the system vulnerabilities. Assuming a rational adversary, who will always choose the action or strategy that rewards highest returns, does not typically map to real-world situations \cite{abbasi2016know}. However, this assumption has been a part of a common staple of approaches that model attacker-defender interactions to compute an \textit{optimal} strategy for the defender, motivated by its practical performance \cite{conitzer2006computing}.

Such adversarial interactions become more interesting and complex when defenders use cyber-deceptive techniques to respond to and thwart attacks. Deception strategies in cybersecurity frameworks, such as installing honeypot configurations to misdirect attackers, have been shown to be effective tools to disrupt attack kill chains and perform attacker reconnaissance \cite{mitrebook15,araujo2014patches,araujo2021software,han2018deception,almeshekah}. Understanding an adversary's behavior can aid cybersecurity defenders to optimally use the available resources to deploy deceptions and mitigate potential threats, while optimizing over the system constraints. Such knowledge can substantially aid a decision-making model, reducing the magnitude of the assumptions a defender must provide about the adversary to make the model operational. 

In this paper, we build on the insight that an embedded honeypot~\cite{araujo16chapter}---a decoy environment that is inlined with genuine service functionality---can be configured in a way that is invisible to attackers while providing the defender with essential knowledge about the attacker's techniques in real operational settings. Moreover, we show that the expected payoffs for the defender may vary when compared to the real-world scenario where the attacker's behaviors may not meet the defender's expectations or prior beliefs.

Our contributions are summarized as follows:
\begin{itemize}
    \item To ground this problem, we created a real-world Capture-The-Flag (CTF) environment hosting three system vulnerabilities, and we conduct studies using human subjects with varying system and attack skill levels who try to capture the corresponding (real) flags. Each of these vulnerabilities are protected by different real and practical mitigation strategies, one of which is a deception-based honey-patch \cite{araujo2014patches}, which misdirects an adversary to an embedded honeypot configuration that yields the attacker a fake, or as we refer to it in this work, a \emph{honeypot flag}.
    \item In parallel to these studies, we model this attacker-defender system interaction as a Markov Game and find its Bayesian Stackelberg Equilibrium. We start with the assumption of inputs to this Markov Game being set by a system expert.
    \item Later, by varying these inputs, especially for cases when the attacker may be trapped in a honeypot configuration, we leverage the statistical results received from the conducted user studies, which further allows us to understand the differences between the obtained equilibria and the empirical setting.
\end{itemize}

The paper is structured as follows: we begin in Section \ref{sec: background} by providing a background on the CTF environment, Markov Games, and the system vulnerabilities with the corresponding mitigations used in this work. We present our hypotheses over the adversary behaviors and explain our user study setup along with the Markov Game modeling in Section \ref{sec:methodology}. Experimental details and results, particularly involving a case-by-case evaluation and discussion on the observations is presented in Section \ref{sec:expts}. We then talk about related work in Section \ref{sec:relatedwork}, with the conclusion discussing future directions for this work in Section \ref{sec:conclusion}.

\section{Background}
\label{sec: background}

In this section, we first present a brief overview of the Capture-The-Flag style setup that we employed to host our user studies. Then, we introduce the real-world vulnerabilities that we used to design the user study test-bed and game-theoretic model evaluations, followed by the defense mechanisms deployed as mitigation strategies. We also describe the Markov Game formalism used for finding an optimal strategy for the defender.

\subsection{Capture-The-Flag Setup}
\label{subsec:ctf}

The primary goal of conducting the user studies is to gather realistic data on attacker behaviors using CTF environments, rather than artificially generating the data based on commonly accepted assumption over adversaries \cite{sengupta2019general}. One way to achieve this is through creating prototype components to run CTF style experiments. We further integrate them into an existing open-source framework known as \emph{the iCTF framework}~\cite{205237,ictf}\footnote{\url{https://github.com/shellphish/ictf-framework}}.
This infrastructure allowed experiments to be run with a sizable number of human subjects to gather enough data for our desired analysis.

The \emph{iCTF framework} is the core framework used for conducting user studies. It is primarily used to host attack-and-defense style CTF competitions every year\footnote{\url{https://shellphish.net/ictf/}}. For the purpose of collecting data for this study, we made several modifications to the existing implementation of the framework. Most of these modifications include deploying defense mechanisms and data collection tools. Since our goal is to simulate real-life scenarios, we choose three vulnerabilities (which are still prominent in current software applications) and develop three corresponding vulnerable applications for this purpose. The vulnerabilities are selected and deployed in a manner that it is possible to exploit them in a reasonable amount of time (which we verified through pilot studies), therefore faithfully representing typical large-scale cyber-attacks. The vulnerabilities selected include command injection and buffer overflow. The vulnerable applications are written in \textit{C} and dockerized to isolate them from the host machine. Also, all modern security mitigations, including Position Independent Executable (PIE), Data Execution Prevention (DEP), and Address Space Layer Randomization (ASLR) are disabled.

\subsection{Vulnerabilities and Exploits}
\label{subsec:vulnerabilities}

We developed three different vulnerable applications. \textit{backup} is the first application which allows users to store and retrieve data that is stored as files on the host system. One of the functions in this application concatenates a string with the user input and passes that string to the C function \texttt{system()}, and the user's input is not sanitized, thus resulting in a command injection vulnerability.

The second application, \textit{sampleak}, allows users to store and retrieve notes which are also stored as files, but unlike the \textit{backup} application, a password is stored in the files, so that the user is required to provide a password when creating a note and needs to enter the correct password when retrieving them. The user input is stored in the application's memory using buffers, but the function \texttt{read()} unintentionally reads in more bytes than the buffer can hold, thus resulting in a buffer overflow vulnerability.

The third vulnerable application is \textit{exploit-market}, which allows users to store, retrieve, and list payloads, which are stored in the memory of the program. The vulnerability in this application is due to buffers being initialized with different sizes in separate functions, so when the function \texttt{strcpy()} is called to copy the contents of the buffer, a carefully crafted payload can result into a buffer overflow vulnerability. Another intentional bug is also placed in the form of a memory disclosure which leaks heap addresses of the string buffers.

\subsection{Defense Strategies and Analysis Tools}
\label{subsec:defenses}

% To test the mitigation strategies for the defender, we then develop defense mechanisms deployed to secure the three vulnerable applications. Two defense mechanisms are selected, specifically \textit{Snort} \textbf{cite}, which is a widely used intrusion detection and prevention system. The other mechanism is \textit{insider} \textbf{cite}, which is a \textit{just in time} (JIT) implementation for honey-patching that works as the deceptive strategy against the attackers.

The defense mitigations are selected for protecting the vulnerable applications. The mitigations include deploying \textit{Snort}, an intrusion detection system on a router machine acting as a gateway between the attacker machine and the defender machine. \textit{Snort} uses a rule-based configuration file for setup, and this rule filter has a list of commonly used \textit{shellcodes} for exploiting various applications running on multiple architectures. We also extended a live-patching framework~\cite{araujo2020insider} to enable cyber-deceptive attack countermeasures. 

To collect valuable attacker and defender information, we further deploy tools on our host machines that include: \textit{tcpdump}, which is a network packet analyzer to capture network traffic for further analysis, and \textit{SysFlow} \cite{taylor2020sysflow,sysflowgh}, an open-source system-call monitoring framework that encodes the representation of system activities into a compact entity-relational format that captures the interactions of processes with system resources, including file and network activity. This provides a richer context for post-exploitation analysis \cite{araujo2021pluggable}.

\subsection{Attack Graph}
\label{subsec:attackgraph}

Attack graphs (AGs) have been established as useful structures to represent exploit possibilities and derive attack behaviors for an adversary \cite{durkota2015optimal,letchford2013optimal}. An attack graph is represented as $\mathcal{G}(\mathcal{V},\mathcal{E})$, where $v \in \mathcal{V}$ denotes vertices or nodes representing the different states the adversary can be in, and $e \in \mathcal{E}$ denotes the edges between these nodes that represent the actions the adversary can take to move one from one state of the exploit to another.

\begin{figure}[tb]
\centering
\includegraphics[width=0.7\textwidth]{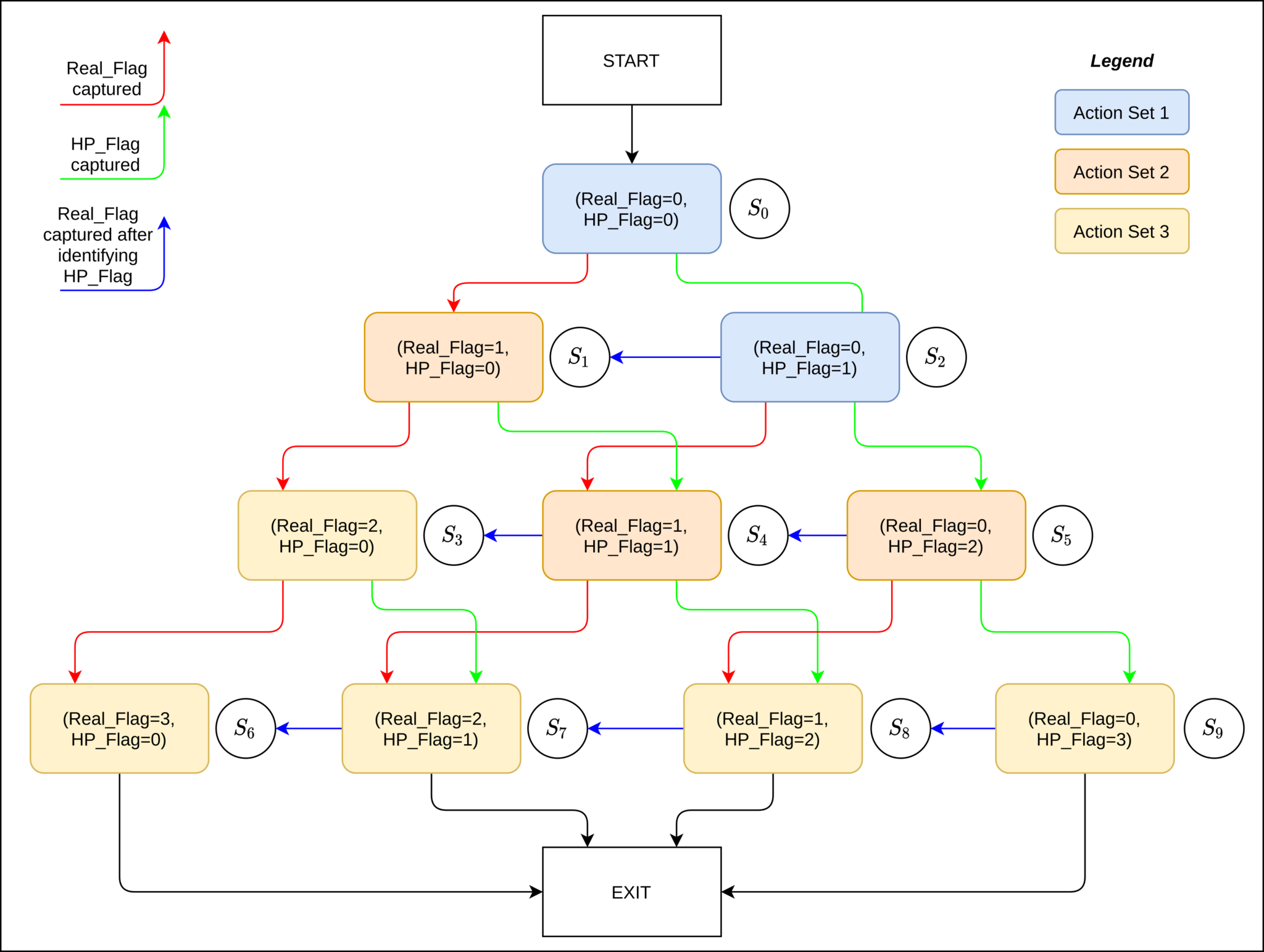}
\caption{A complete representation of the attacker's possible attack graphs in the current game setup.} 
\label{fig:attack_graph}
\end{figure}

Figure \ref{fig:attack_graph} is an example of an attack graph for an attacker trying to exploit the vulnerabilities present in the environment with the possibility of one or more of them being honey-patched, i.e., deceiving and misdirecting the attacker into a honeypot configuration where the system defender can extract useful insights about attacker behavior.

As noted in \cite{attack_graphs_review}, an attack graph can be represented as the tuple $\mathcal{G} = (S, \tau, S_{0}, S_{S}, L, EX)$, where:

\begin{itemize}
    \item $S$ is the finite set of states or nodes, in our case a total of 10, on one of which the attacker will be present during the exploit,
    \item $\tau \subseteq S \times S$ represents the transition function which defines the probabilities of the attacker taking an available action in a state and reaching another state,
    \item $S_{0} \subseteq S$ represents the set of initial states, and in our case, it is the state where the attacker has captured 0 real or honeypot flags,
    \item $S_{S} \subseteq S$ represents the set of final states (success or failure) for the attacker, and in our case, states where the attacker has tried exploiting each of the three vulnerabilities and has either captured the real flag or the honeypot flag,
    \item $L$ represents the atomic propositions used for labeling these states or nodes, which in our case correspond to the number of real flags and honeypot flags captured by the attacker,
    \item and, finally, $EX$ represents a finite set of actions, such as shown by different colored nodes representing different sets of actions available to the attacker in our attack graph. We talk about the different action sets corresponding to each of the states separately in Section \ref{subsec:game_modeling}.
\end{itemize}

\subsection{Markov Game}
\label{subsec:markovgame}

We define a Markov Game, and the associated notations, between the attacker ($\mathcal{A}$) and defender ($\mathcal{D}$) as:

\begin{itemize}
    \item a set of states $S$ representing the collection of states,
    \item action set $A:(A_{\mathcal{A}} \times A_{\mathcal{D}})$ which comprises of the cross-product between the action sets of the attacker ($A_{\mathcal{A}}$) and that of the defender ($A_{\mathcal{D}}$),
    \item $T$ represents the transition probability matrix from a state $s_{i} \in S$ to $s_{j} \in S$ when the attacker takes an action $a_{\mathcal{A}} \in A_{\mathcal{A}}$ and the defender takes an action $a_{\mathcal{D}} \in A_{\mathcal{D}}$,
    \item $U(s \in S, a \in A)$ denotes the utility or the rewards received by the player ($\mathcal{A}$ or $\mathcal{D}$) in state $s$ when action $a$ is taken, and
    \item to take discounted future rewards into consideration, we define $\gamma:[0,1)$ for both the players $\mathcal{A}$ and $\mathcal{D}$.
\end{itemize}

Basak et al.~\cite{varying_gamma} highlight that it is not straightforward to make the assumption that the $\gamma$ would be the same for both $\mathcal{A}$ and $\mathcal{D}$. Following the limitations or the absence of a formal study as shown in \cite{sengupta2019general}, we take a similar approach and assume for the time being that $\gamma = \gamma_{\mathcal{A}} = \gamma_{\mathcal{D}}$.

For the zero-sum game that we assume in this work, an optimal policy for the defender's strategy can be computed \cite{littman1994markov}, and can be updated to induce a min-max strategy for the two players \cite{sengupta2019general}, by calculating the $\mathcal{Q}$-value or the expected return for an actor ($\mathcal{A}$ or $\mathcal{D}$) in state $s \in S$ and taking action $a_{i}$ for $i \in (\mathcal{A}, \mathcal{D})$ as:

\begin{equation}
\label{eq:q_value}
    \mathcal{Q}(s, a_{\mathcal{D}}, a_\mathcal{A}) = U(s, a_{\mathcal{D}}, a_\mathcal{A}) + \gamma \sum_{s'}T(s, a_{\mathcal{D}}, a_\mathcal{A}, s')\mathcal{V}(s')
\end{equation}

\noindent where the defender takes action $a_{\mathcal{D}}$ against the attacker's action $a_{\mathcal{A}}$ and reaches another state $s'$. The value-function for the defender's mixed policy $\pi(s)$ for the probability $\pi_{a_{\mathcal{D}}}$ of choosing action $a_{\mathcal{D}}$ is defined as:

\begin{equation}
\label{eq:value_function}
    \mathcal{V}(s) = \max_{\pi(s)}\min_{a_{\mathcal{A}}}\sum_{a_{\mathcal{D}}}Q(s, a_{\mathcal{D}}, a_\mathcal{A})\pi_{a_{\mathcal{D}}}
\end{equation}

We use this to compute the optimal mixed strategy for our defender for the different experimental settings, as explained in more detail in Section \ref{sec:expts}. Given the components of our Markov Game for a fixed set of utilities, which are dependent on the exploitable vulnerabilities and available mitigations used, the transition probability matrix $T(s, a_{\mathcal{D}}, a_\mathcal{A}, s')$ may vary based on the threat level posed by an adversary. This motivates us to test our Markov Game modeling by varying these input parameters, and we discuss these variations further in detail in Section \ref{subsec:game_modeling}.

\section{Methodology}
\label{sec:methodology}

In the current setup, we assumed that the system contains 3 vulnerabilities, $exploit_{i}$  for $i \in [1,3]$. One or more of these vulnerabilities may be honey-patched by the system defender. Honey-patches~\cite{araujo2014patches} are software security patches that are modified to avoid alerting adversaries when their exploit attempts fail. Instead of blocking the attempted exploit, the honey-patch transparently redirects the attacker session to an isolated honeypot environment. Adversaries attempting to exploit a honey-patched vulnerability
observe software responses that resemble unpatched software, even though the vulnerability is actually patched. This allows defenders to observe attack actions until the deception is uncovered.

Figure~\ref{fig:attack_graph} shows a scenario where all three vulnerabilities are honey-patched. Note that the attack graph for a case where one or more vulnerabilities are not honey-patched will be a special case of the attack graph shown in the figure, with those particular vulnerabilities leading to a real flag, and not to a 
%honey-patched
honeypot flag.

In the beginning of the game, the attacker has the option to exploit one of the vulnerabilities, denoted by starting state $(real\_flag=0, hp\_flag=0)$. To illustrate how the game advances, assuming that the initial target is the honey-patched vulnerability $exploit_1$, the resulting attacker state would be $(real\_flag=0, hp\_flag=1)$, denotation the state where the attacker is trapped into a decoy and fails to capture the real flag. 
%Based on the two-hypothesis listed down below, 
The attacker then progresses through the game and eventually reaches a terminal state $(real\_flag=i, hp\_flag=j)$, where $i \in [0,3]$ \& $j \in [0,3]$ and $i+j \leq 3$, since an attacker can obtain at most either of the two flags for each exploit, or none at all. Section~\ref{sub: hypothesis} details our hypotheses about the dynamics of the game.

\subsection{Hypotheses}
\label{sub: hypothesis}

We design our study to test two different sets of strategies that can be adopted by an attacker once trapped into a honeypot configuration. We also state here the assumptions behind each hypothesis that we test.

\begin{itemize}[leftmargin=20pt]
    \item[\textbf{H1}] Once trapped in a honeypot environment, the attacker chooses to \emph{continue} with the existing strategy to exploit the remaining vulnerabilities.
    \item[\textbf{H2}] Once trapped in a honeypot environment, the attacker chooses to \emph{change} the current strategy to exploit the remaining vulnerabilities.
\end{itemize}

\noindent\textbf{Hypothesis 1} assumes that the attacker did not discover the deception, i.e., a state where a honeypot flag was obtained. Hence, the attacker chooses to 
to continue with the existing strategy without worrying about subsequent traps.

\vspace{6pt}\noindent\textbf{Hypothesis 2}, on the other hand, assumes that the attacker becomes aware of the new state, i.e., a state where a honeypot flag is discovered. Hence, the attacker chooses to change the current strategy (i.e., attempts to escape the honeypot environment). This would allow the attacker to retry exploiting the honey-patched vulnerability while avoiding getting trapped in the honeypot. In case the attempt to escape the particular honeypot configuration is unsuccessful, the attacker awareness of the deception will influence the attack strategy to be more cautious and observant of subsequent honey-patches.

\vspace{6pt}Next, we discuss the user study that allows us to track the varying attack behaviors generated by each of the adversaries, followed by our Markov Game modeling used for running the experiments.

\subsection{User Study}

The \textit{ictf framework} is intended to be used for hosting multi-team (over 100 have been accomplished) attack-defense style CTFs. In our user studies, the goal is to have a human participant assuming the role of an attacker and several security defense mechanisms assuming the role of defenders. The attacker's machine has identical copies of the vulnerable applications to enable attackers to analyze the application, identify vulnerabilities, and develop working exploits. On the defender's machines, the vulnerable applications are deployed with either one of the security mitigations or no mitigation.

A pre-generated string called the \textit{flag} is placed inside the root directory of the docker containers running the vulnerable applications. The goal of the attacker is to successfully exploit the vulnerabilities and read the flag, which is only accomplished by successfully exploiting the vulnerable application. 

\begin{figure}[tb]
\centering
\frame{\includegraphics[width=0.5\textwidth]{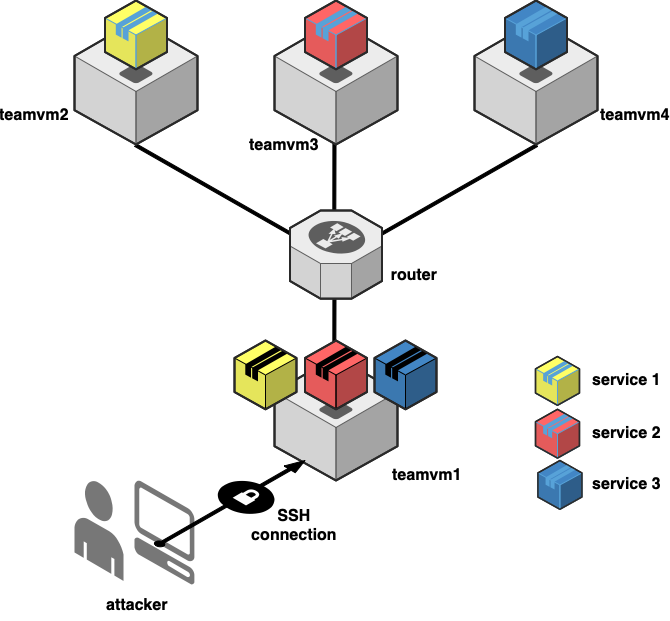}}
\caption{iCTF infrastructure setup for running experiments.} 
\label{fig:ictf_setup}
\end{figure}

% \begin{figure}[tb]
% \centering
% \begin{subfigure}{.4\textwidth}
%   \centering
%   \includegraphics[width=\linewidth]{images/ictf_setup.png}
%   \caption{iCTF infrastructure setup.}
%   \label{fig:ictf_setup}
% \end{subfigure}%
% \begin{subfigure}{.4\textwidth}
%   \centering
%   \includegraphics[width=\linewidth]{images/iCTF.png}
%   \caption{iCTF experimental setup.}
%   \label{fig:sample_experiment}
% \end{subfigure}
% \caption{User study setup for conducting experimental studies using iCTF.}
% \label{fig:ictf_images}
% \end{figure}

Figure \ref{fig:ictf_setup} shows the infrastructure setup. The router serves as a gateway between all of the team virtual machines (VMs). The attacker participant is granted SSH access to the team-VM1 machine. They can only interact with the vulnerable services through the team-VM1 machine. All team-VMs run the same environment (Ubuntu 18.04) and have copies of docker images for all vulnerable applications, which will be referred to as services. As a result, the attacker has access to copies of all the services used in the experiment in order to analyze and develop exploits for them.

\begin{figure}[tb]
\centering
\frame{\includegraphics[width=0.5\textwidth]{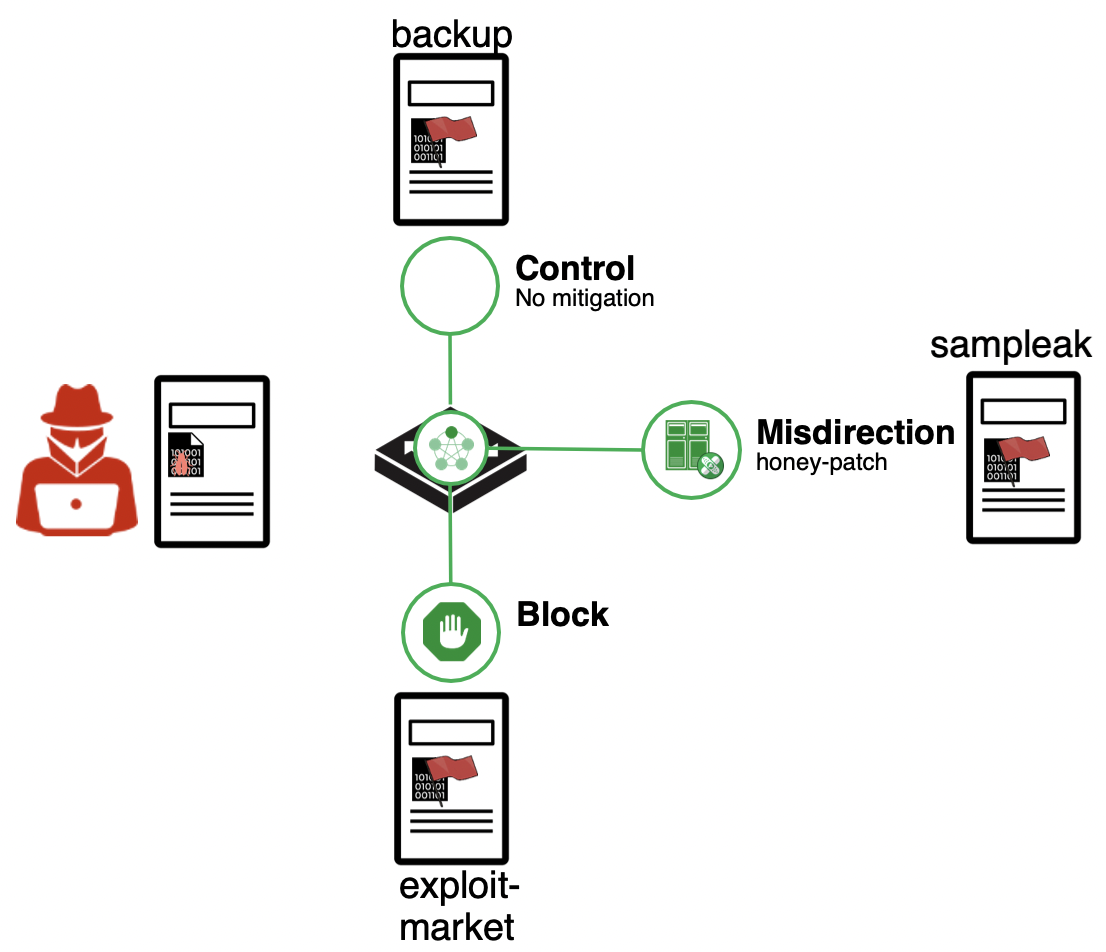}}
\caption{The experimental setup used for conducting the iCTF user studies.}
\label{fig:sample_experiment}
\end{figure}

Once the infrastructure is created, a random setup of services and defenses is selected. The service-defense relationship is one-to-one. Figure \ref{fig:sample_experiment} shows one randomly selected setup wherein \textit{backup} has no defense mitigation, \textit{sampleak} has \textit{honey-patching} as a mitigation and \textit{exploit-market} has a \textit{Snort} filter as a mitigation. One participant at a time assumes the role of an attacker and we describe that their goal is to identify vulnerabilities in the vulnerable service and write an exploit for the vulnerability that reads the flag. The participant is given sufficient time to work on each service. Whether they successfully exploit it or run out of time, they try to exploit one service at a time, in the consistent order of: \textit{backup}, \textit{sampleak}, \textit{exploit-market}. Once the experiment concludes, the data is collected and stored for further analysis.

\subsection{Markov Game Modeling}
\label{subsec:game_modeling}

The different stages of the attack graph in Figure \ref{fig:attack_graph} represent the diverse scenarios that an attacker might be in, when trying to exploit the different vulnerabilities present on the defender's system. In each of these unique stages, the attacker has a set of actions available to execute on the system. Intuitively, as the game progresses, the number of possible actions for both the attacker and the defender decreases, which limits their control over the game. Hence, we formulate the model which can recommend the action probabilities for the defender at each of these stages. We further analyze what varying costs the defender incurs when taking each of these actions and how the future expectations of the defender’s actions can influence the adopted mitigation strategies.

\subsubsection{State space} $S$ for the Markov Game $\mathcal{M}$ is defined by the nodes of the attack graph.
%as shown in Figure \ref{fig:attack_graph}. 
The attacker enters the system to begin exploits without any real or honeypot flags, represented by state $S_{0}$. Similarly, after attempting exploits on each of the three vulnerabilities, the attacker may end up with a set of real and honeypot flags and be end in one of the terminal states from $S_{6}$ to $S_{9}$, after which the attacker exits the system. The blue, orange, and yellow colors in these nodes represent the action sets available to the agent in each of these states, and we explain these action states in more detail next.

\subsubsection{Action space} $A$ for the Markov Game $\mathcal{M}$ comprises of the action sets $A_{\mathcal{A}} \times A_{\mathcal{D}}$, i.e., of both the players. Intuitively, as the game progresses and the number of vulnerabilities remaining to exploit for the attacker reduce, the possible action set also minimizes for the attacker, and also for the defender given the naturally assumed resource constraints. Note that the defender takes actions before the game begins and assigns the desired set of mitigations on one or more of the existing system vulnerabilities.

%As shown in the attack graph of Figure \ref{fig:attack_graph}, 
In the example attack graph shown in Figure~\ref{fig:attack_graph}, Action Set 1 comprises of four actions for the attacker: \textit{do nothing}, \textit{exploit vulnerability 1}, \textit{exploit vulnerability 2} or \textit{exploit vulnerability 3}. For the defender, the four actions available are: \textit{no mitigation}, \textit{honey-patch vulnerability 1}, \textit{honey-patch vulnerability 2} or \textit{honey-patch vulnerability 3}.

Further in the game for Action Set 2, the attacker has three actions remaining as the attacker cannot return to exploiting the first vulnerability. Similarly, the defender has all the actions available except deploying honey-patch for the first vulnerability for which an exploit was attempted.

Lastly, for Action Set 3, the attacker has two actions remaining: \textit{do nothing}, and \textit{exploit vulnerability 3}. Similarly, the defender has two actions remaining: \textit{no mitigation}, and \textit{honey-patch vulnerability 3}.

\begin{table}[tb]
\centering
\caption{Utility Matrix for Action Set 1}
\label{tab:actionset1}
\begin{tabular}{cccccc}
                                                                                          & \multicolumn{5}{c}{\textbf{Defender's actions ($A_{\mathcal{D}}$)}}                                                                                           \\
                                                                                          &                             & no\_mon                   & hp\_1                     & hp\_2                     & hp\_3                     \\ \cline{3-6} 
                                                                                          & \multicolumn{1}{c|}{no\_op} & \multicolumn{1}{c|}{0}    & \multicolumn{1}{c|}{-3}   & \multicolumn{1}{c|}{-3}   & \multicolumn{1}{c|}{-3}   \\ \cline{3-6} 
\multirow{2}{*}{\begin{tabular}[c]{@{}c@{}}\textbf{Attacker's} \\ \textbf{actions ($A_{\mathcal{A}}$)}\end{tabular}} & \multicolumn{1}{c|}{exp\_1} & \multicolumn{1}{c|}{-5.9} & \multicolumn{1}{c|}{2.9}  & \multicolumn{1}{c|}{-8.9} & \multicolumn{1}{c|}{-8.9} \\ \cline{3-6} 
                                                                                          & \multicolumn{1}{c|}{exp\_2} & \multicolumn{1}{c|}{-5.9} & \multicolumn{1}{c|}{-8.9} & \multicolumn{1}{c|}{2.9}  & \multicolumn{1}{c|}{-8.9} \\ \cline{3-6} 
                                                                                          & \multicolumn{1}{c|}{exp\_3} & \multicolumn{1}{c|}{-5.9} & \multicolumn{1}{c|}{-8.9} & \multicolumn{1}{c|}{-8.9} & \multicolumn{1}{c|}{2.9}  \\ \cline{3-6} 
\end{tabular}
\end{table}

\subsubsection{Utility Matrices:} $A_1$ for states $S_0$ and $S_2$ comprises of the actions, along with the corresponding utilities shown in Table \ref{tab:actionset1}. Similarly, Table \ref{tab:actionset2} and Table \ref{tab:actionset3} represent the actions for both the attacker and the defender for states $S_1$, $S_4$, and $S_5$, and, states $S_3$, $S_6$-$S_9$ respectively. By running some initial experimental studies comparing the payoffs received, we conclude that honey-patching offers the highest returns to the defender against any vulnerability exploit, as compared to deploying no mitigation or using \textit{Snort}. We further analyze the cases where the defender has an option of honey-patching one or more vulnerabilities present in the system. Hence, Tables \ref{tab:actionset1}, \ref{tab:actionset2} and \ref{tab:actionset3} only comprise of the honey-patching actions for the defender, and the scores for which have been computed using the scoring system of CVSS v3.1\footnote{\url{https://www.first.org/cvss/v3.1/specification-document}}. These CVSS scores determine the exploitability and severity of the three exploits used in this study.

\begin{table}[tb]
\centering
\caption{Utility Matrix for Action Set 2}
\label{tab:actionset2}
\begin{tabular}{ccccc}
                                                                                         &                             & \multicolumn{3}{c}{\begin{tabular}[c]{@{}c@{}}\textbf{Defender's actions} \\ \textbf{($A_{\mathcal{D}}$)}\end{tabular}} \\
                                                                                         &                             & no\_mon                       & hp\_2                         & hp\_3                        \\ \cline{3-5} 
                                                                                         & \multicolumn{1}{c|}{no\_op} & \multicolumn{1}{c|}{0}        & \multicolumn{1}{c|}{-3}       & \multicolumn{1}{c|}{-3}      \\ \cline{3-5} 
\multirow{2}{*}{\begin{tabular}[c]{@{}c@{}}\textbf{Attacker's} \\ \textbf{actions($A_{\mathcal{A}}$)}\end{tabular}} & \multicolumn{1}{c|}{exp\_2} & \multicolumn{1}{c|}{-5.9}     & \multicolumn{1}{c|}{2.9}      & \multicolumn{1}{c|}{-8.9}    \\ \cline{3-5} 
                                                                                         & \multicolumn{1}{c|}{exp\_3} & \multicolumn{1}{c|}{-5.9}     & \multicolumn{1}{c|}{-8.9}     & \multicolumn{1}{c|}{2.9}     \\ \cline{3-5} 
\end{tabular}
\end{table}

\begin{table}[tb]
\centering
\caption{Utility Matrix for Action Set 3}
\label{tab:actionset3}
\begin{tabular}{cccc}
                                                                                         &                             & \multicolumn{2}{c}{\begin{tabular}[c]{@{}c@{}}\textbf{Defender's actions} \\ \textbf{($A_{\mathcal{D}}$)}\end{tabular}} \\
                                                                                         &                             & no\_mon                                       & hp\_3                                        \\ \cline{3-4} 
\multirow{2}{*}{\begin{tabular}[c]{@{}c@{}}\textbf{Attacker's} \\ \textbf{actions  ($A_{\mathcal{A}}$)}\end{tabular}} & \multicolumn{1}{c|}{no\_op} & \multicolumn{1}{c|}{0}                        & \multicolumn{1}{c|}{-3}                      \\ \cline{3-4} 
                                                                                         & \multicolumn{1}{c|}{exp\_3} & \multicolumn{1}{c|}{-5.9}                     & \multicolumn{1}{c|}{2.9}                     \\ \cline{3-4} 
\end{tabular}
\end{table}

\subsubsection{Transition Matrices} vary based on the modeling of the game. We model three different scenarios to evaluate the results from our user studies. We begin with the assumption that the system defender either does not have any prior knowledge on the threat level posed by the adversary, in which case the transition probabilities can either be set randomly with non-zero probabilities assigned to all feasible state transitions, or, as an expert with the knowledge of the state transitions for any attacker interacting with the environment. The transition matrix for the \textbf{Naive Model} thus comprises of preset probabilities to the different possible transitions. This model can also assume that the defender incurs an equal cost of deploying a honey-patch for each of the three vulnerabilities. Later, we also relax this assumption and assign relatively higher costs to mitigations which are more difficult to deploy by the defender on the system. We randomly generate probabilities for all feasible transitions for the former case, and for the latter, the different probabilities associated with the model are shown in Table \ref{tab:naive_model}.

\begin{table}[tb]
\centering
\caption{Transition probabilities set by system expert for the Naive Model.}
\label{tab:naive_model}
\begin{tabular}{|c|c|} 
\hline
\textbf{Observation}                                                                              & \textbf{Probability}  \\ 
\hline
\begin{tabular}[c]{@{}c@{}}\textbf{Real flag captured}\\\textbf{if no honeypot}\end{tabular}      & 0.75                  \\ 
\hline
\begin{tabular}[c]{@{}c@{}}\textbf{Real flag captured}\\\textbf{if honeypot present}\end{tabular} & 0.4                   \\ 
\hline
\textbf{Trapped in honeypot}                                                                      & 0.4                   \\
\hline
\end{tabular}
\end{table}

Note, that the probability of not capturing any flag for any vulnerability is 0.2 in the case where honeypot is present, and 0.25 when it is not deployed.

For the third model which is the \textbf{Updated Model with Tuned Parameters} using the data collected through the successful iCTF user studies and observations drawn from the attacker participants' behavior, we update the transition matrix for the model to better emulate the real-life scenario. The different probability inputs associated with this model are shown in Table \ref{tab:tuned_param_model}.

\begin{table}[tb]
\centering
\caption{Transition probabilities derived from iCTF user studies' statistics.}
\label{tab:tuned_param_model}
\begin{tabular}{|c|c|c|}
\hline
\textbf{Observation}                                                         & \textbf{Vulnerable Application} & \textbf{Probability} \\ \hline
\multirow{3}{*}{\begin{tabular}[c]{@{}c@{}}\textbf{Real flag captured} \\ \textbf{if no honeypot}\end{tabular}} & \textit{backup}                 & 1.0                                   \\ \cline{2-3} 
                                                                                              & \textit{sampleak}               & 0.43                                  \\ \cline{2-3} 
                                                                                              & \textit{exploit-market}         & 0.4                                   \\ \hline
\multirow{3}{*}{\textbf{Trapped in honeypot}}                                                          & \textit{backup}                 & 1.0                                   \\ \cline{2-3} 
                                                                                              & \textit{sampleak}               & 0.5                                   \\ \cline{2-3} 
                                                                                              & \textit{exploit-market}         & 0.6                                   \\ \hline
\end{tabular}
\end{table}

We also remove the assumption that the defender incurs equal cost of deploying the honeypot configuration as a mitigation, since configuring a honeypot is harder for a stronger suite of vulnerabilities that may be targeted by a system adversary. For all the above-mentioned modeling scenarios, we vary the costs of honey-patching these vulnerabilities accordingly and thus have a uniform and a non-uniform mitigation cost scenario for all three the models. In the non-uniform variants, the minimum cost of 1 is associated with honey-patching \textit{backup} and the maximum cost of 3 is associated with honey-patching \textit{exploit-market}. On the other hand, for the uniform cost variants, a cost of 3 is associated with deploying a honey-patch against each of the three vulnerabilities.

\section{Experimental Evaluation and Results}
\label{sec:expts}

\subsection{iCTF User Studies}
\label{subsec:user_studies_results}

After running six pilot studies, and making necessary infrastructure changes to gather data on the adversary's behavior, we conducted 18 user studies. Participants were recruited from a pool of people known to have prerequisite skill-set (vulnerability analysis and software exploitation) and were rewarded \$50 USD for their participation. Prior to this, we obtained an IRB approval from 
%Arizona State University 
our institution for conducting the user studies.

Table  \ref{tab:summary} shows the summary of these user studies. The first column represents each service used in the experiments. The second column represents the total number of experiments conducted, and the third column represents the number of times the participant timed out without being able exploit a particular service. The last column is broken down by the defense mechanism deployed, the numerator represents successful exploitation of the service, and the denominator represents the total number of times the service is deployed with the particular defense.

\begin{table}[tb]
\centering
\caption{Summary Of iCTF Experiments.}
\label{tab:summary}
\resizebox{\textwidth}{!}{
\begin{tabular}{|c|c|c|c|c|c|c|} 
\cline{4-7}
\multicolumn{1}{c}{}          & \multicolumn{1}{c}{}          &                    & \multicolumn{4}{c|}{\textbf{Flag Captured w/ Defense Mechanism}}                       \\ 
\cline{4-7}
\multicolumn{1}{c}{\textbf{}} & \multicolumn{1}{c}{\textbf{}} &                    & \textbf{None} & \textbf{Snort}          & \multicolumn{2}{c|}{\textbf{Honeypatch}}     \\ 
\hline
\textbf{Challenge}            & \textbf{Total Experiments}    & \textbf{Timed Out} & \multicolumn{2}{c|}{\textbf{Real Flag}} & \textbf{Real Flag} & \textbf{Honeypot Flag}  \\ 
\hline
\textbf{backup}               & 18                            & 0                  & 6/6           & 6/6                     & 0/6                & 6/6                     \\ 
\hline
\textbf{sampleak}             & 18                            & 9                  & 4/6           & 2/5                     & 0/7                & 3/7                     \\ 
\hline
\textbf{exploit-market}       & 15                            & 6                  & 3/5           & 3/5                     & 0/5                & 3/5                     \\
\hline
\end{tabular}
}
\end{table}

\subsection{Markov Game Strategy Evaluation}
\label{subsec:markov_game_results}

We first discuss the performance of our zero-sum Markov Game model against Uniform Random Strategy (URS), which has been shown to be effective in similar attacker-defender settings \cite{Zhuang14,Taguinod15,Winterrose20}, and Min-Max Pure (MMP) Strategy. The utility payoffs in our results correspond to the payoffs of the defender. As it has been previously shown~\cite{sengupta2019general}, Optimal Mixed Strategy outperforms the other two algorithms with respect to returns gained by the system defender. By testing the three algorithms for the \textbf{Naive Model} with both randomly set and expertly assumed set of transition probabilities with uniform costs of deploying honeypots, we establish a baseline that would later help us understand how additional knowledge of the attacker that the system defender gained, can improve upon the decision making model. 

With reference to the state space represented in Figure \ref{fig:attack_graph}, we show the results for state $S_2$ where the attacker is trapped in the first honeypot, state $S_5$ (similar to $S_4$) where the attacker is trapped in the second honeypot, and finally state $S_8$ (similar to state $S_7$ and state $S_9$) where the attacker is trapped in the honeypot set for the third vulnerability. We are more interested to analyze these states where deception is successful for the defender, and thus we would want to compare the usefulness gained by the user-studies later, particularly for these states.

\begin{figure}[tb]
\centering
\begin{subfigure}{.3\textwidth}
  \centering
  \includegraphics[width=\linewidth]{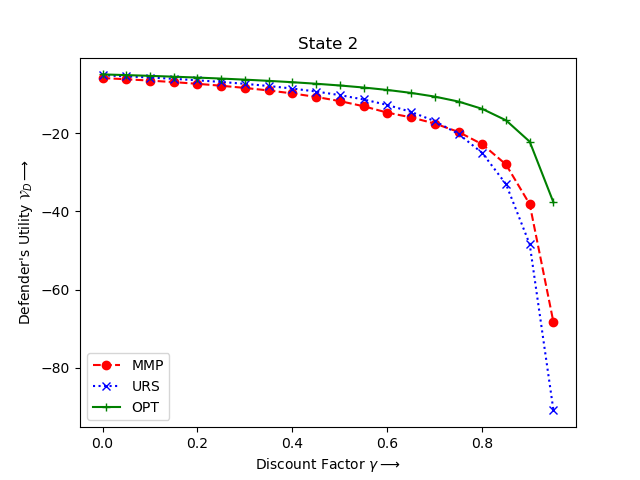}
  \caption{Returns for state $S_2$.}
  \label{fig:random_uniform_s2}
\end{subfigure}%
\begin{subfigure}{.3\textwidth}
  \centering
  \includegraphics[width=\linewidth]{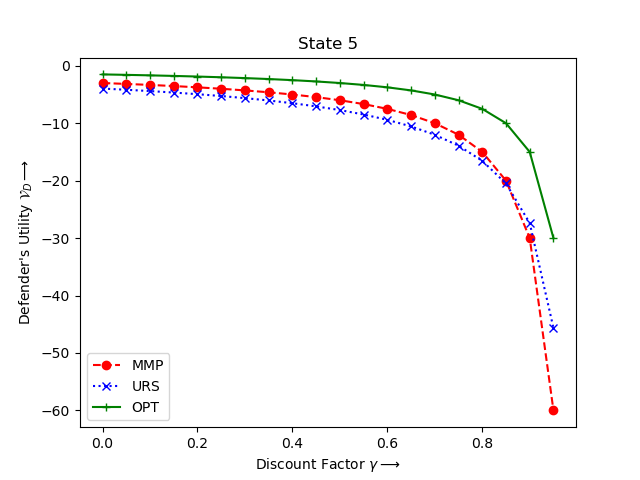}
  \caption{Returns for state $S_5$.}
  \label{fig:random_uniform_s5}
\end{subfigure}
\begin{subfigure}{.3\textwidth}
  \centering
  \includegraphics[width=\linewidth]{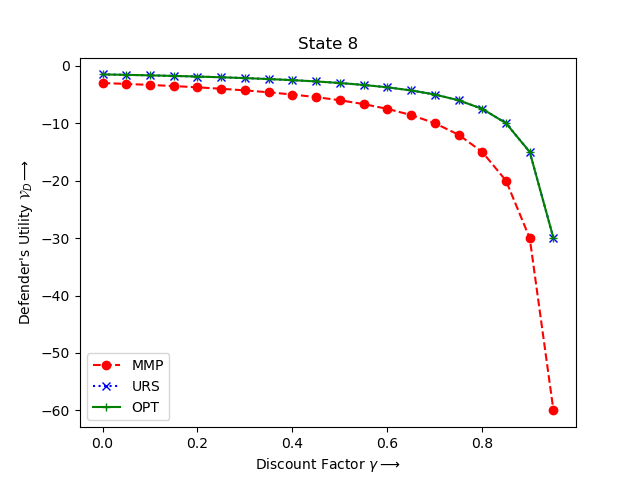}
  \caption{Returns for state $S_8$.}
  \label{fig:random_uniform_s8}
\end{subfigure}
\caption{Defender's payoffs for Naive Model - randomly set transition probabilities and uniform mitigation deployment costs.}
\label{fig:naive_random_uniform}
\end{figure}

\begin{figure}[tb]
\centering
\begin{subfigure}{.3\textwidth}
  \centering
  \includegraphics[width=\linewidth]{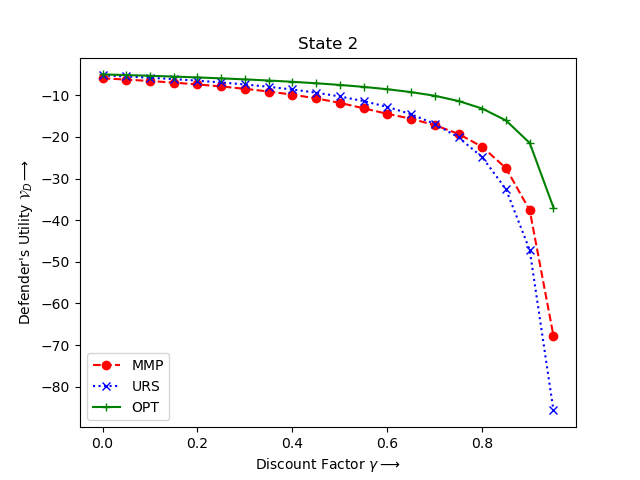}
  \caption{Returns for state $S_2$.}
  \label{fig:expert_uniform_s2}
\end{subfigure}%
\begin{subfigure}{.3\textwidth}
  \centering
  \includegraphics[width=\linewidth]{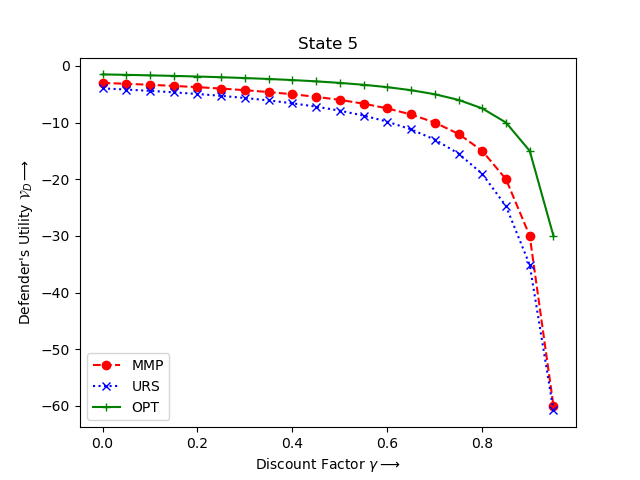}
  \caption{Returns for state $S_5$.}
  \label{fig:expert_uniform_s5}
\end{subfigure}
\begin{subfigure}{.3\textwidth}
  \centering
  \includegraphics[width=\linewidth]{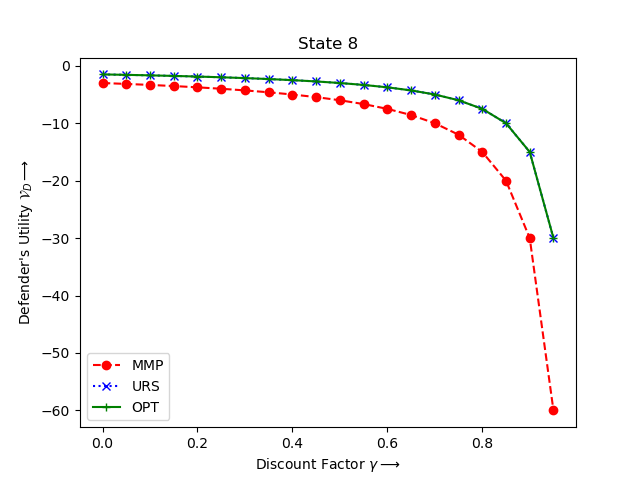}
  \caption{Returns for state $S_8$.}
  \label{fig:expert_uniform_s8}
\end{subfigure}
\caption{Defender's payoffs for Naive Model - system expert set transition probabilities and uniform mitigation deployment costs.}
\label{fig:naive_expert_uniform}
\end{figure}

\textcolor{black}{Figures \ref{fig:naive_random_uniform} and \ref{fig:naive_expert_uniform} show that Optimal Mixed Strategy (OPT) outperforms MMP and URS for states seen early (Figures \ref{fig:random_uniform_s2}, \ref{fig:expert_uniform_s2}) and in the middle (Figures \ref{fig:random_uniform_s5}, \ref{fig:expert_uniform_s5}) of the game, and returns payoffs for the defender equal to URS in the later stages (Figures \ref{fig:random_uniform_s8}, \ref{fig:expert_uniform_s8}) of the game where both URS and OPT output the same distribution over the remaining two actions for the defender.} 

We further apply the Optimal Mixed Strategy to the updated model scenario that we discussed in Section \ref{sec:methodology}. We only focus on the Optimal Mixed Strategy to compare the Stackelberg equilibrium obtained under the following case studies. We divide each case study using the discount factor $\gamma \in [0,1)$ to understand the variation in the model's decision making as the weightage for future exploit mitigations varies for the system defender, and particularly discuss the two corner cases of $\gamma = 0$ and $\gamma=0.95$. We compare the equilibria obtained among the three model variations under uniform costs of deploying mitigations, i.e., \textbf{Naive Model} with randomly set transition probabilities, \textbf{Naive Model} with expertly assumed transition probabilities, and \textbf{Updated Model with Tuned Parameters}. We carry out a similar comparison among the three models for non-uniform costs later.

\begin{figure}[tb]
\centering
\begin{subfigure}{.3\textwidth}
  \centering
  \includegraphics[width=\linewidth]{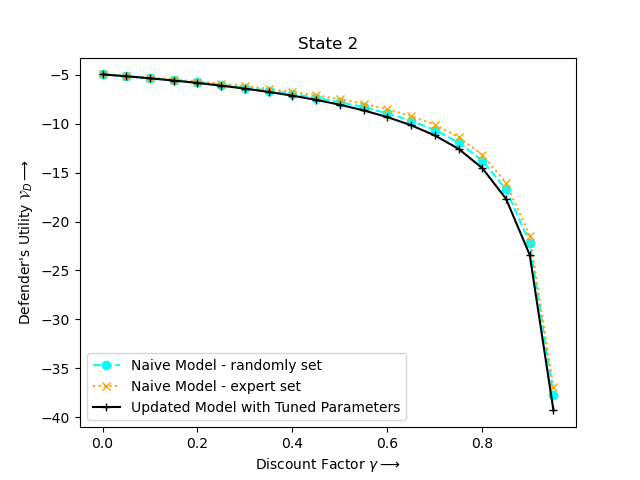}
  \caption{Returns for state $S_2$.}
  \label{fig:uniform_model_s2}
\end{subfigure}%
\begin{subfigure}{.3\textwidth}
  \centering
  \includegraphics[width=\linewidth]{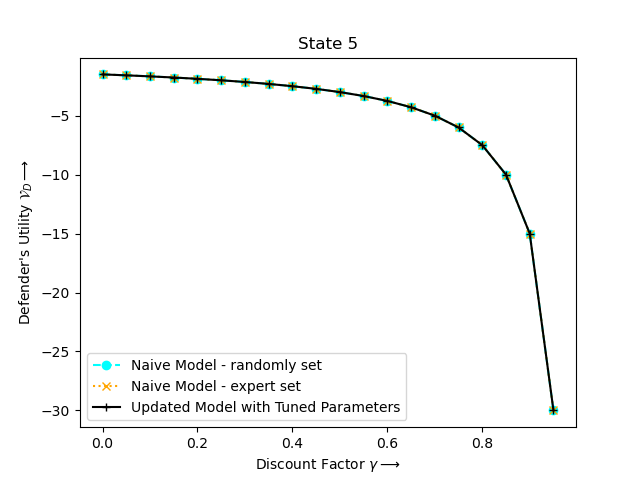}
  \caption{Returns for state $S_5$.}
  \label{fig:uniform_model_s5}
\end{subfigure}
\begin{subfigure}{.3\textwidth}
  \centering
  \includegraphics[width=\linewidth]{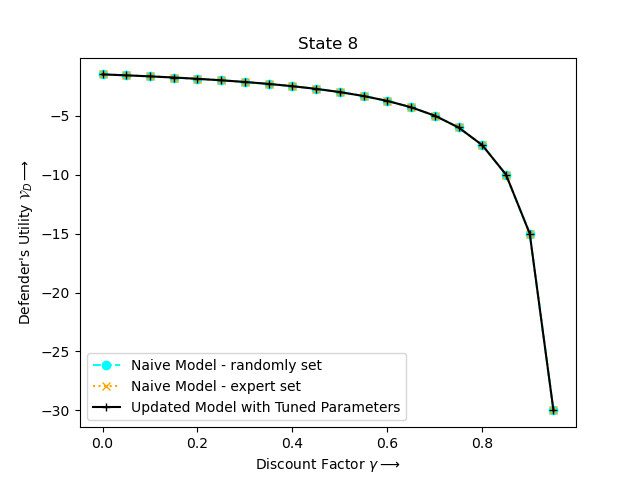}
  \caption{Returns for state $S_8$.}
  \label{fig:uniform_model_s8}
\end{subfigure}
\caption{Defender's payoffs compared for the three models using uniform mitigation deployment costs.}
\label{fig:uniform_model}
\end{figure}

\begin{figure}[tb]
\centering
\begin{subfigure}{.3\textwidth}
  \centering
  \includegraphics[width=\linewidth]{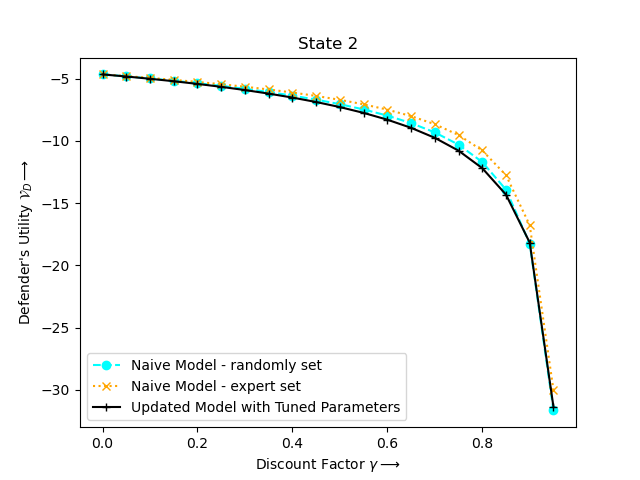}
  \caption{Returns for state $S_2$.}
  \label{fig:non_uniform_model_s2}
\end{subfigure}%
\begin{subfigure}{.3\textwidth}
  \centering
  \includegraphics[width=\linewidth]{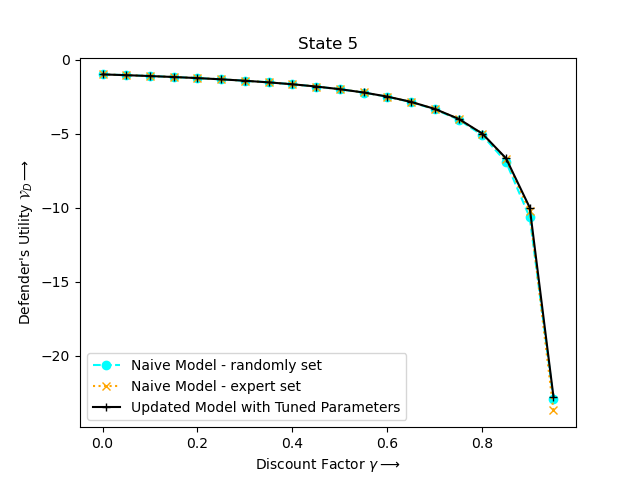}
  \caption{Returns for state $S_5$.}
  \label{fig:non_uniform_model_s5}
\end{subfigure}
\begin{subfigure}{.3\textwidth}
  \centering
  \includegraphics[width=\linewidth]{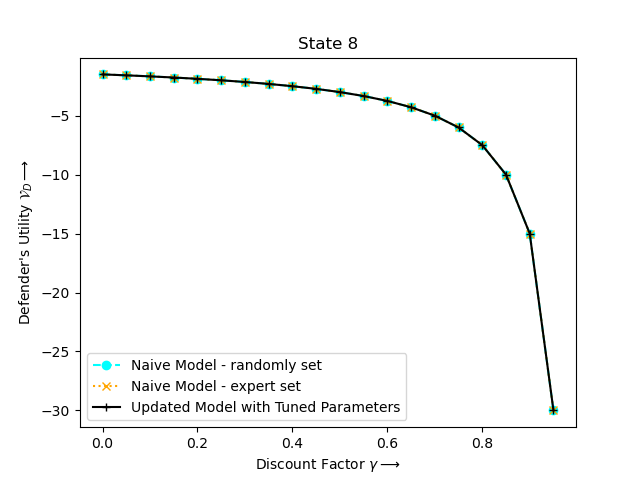}
  \caption{Returns for state $S_8$.}
  \label{fig:non_uniform_model_s8}
\end{subfigure}
\caption{Defender's payoffs compared for the three models using non-uniform mitigation deployment costs.}
\label{fig:non_uniform_model}
\end{figure}

\begin{table}[tb]
\centering
\caption{Comparing the three modeling scenarios for Case Study 1}
\label{tab:casestudy1}
\begin{tabular}{|c|c|c|c|c|} 
\hline
\multirow{2}{*}{\textbf{Case Study 1}}                                                                           & \multirow{2}{*}{\textbf{$\gamma$}} & \multicolumn{3}{c|}{\textbf{Probability of honey-patching}}    \\ 
\cline{3-5}
                                                                                                                 &                                    & \textit{backup} & \textit{sampleak} & \textit{exploit-market}  \\ 
\hline
\multirow{2}{*}{\begin{tabular}[c]{@{}c@{}}\textbf{Naive Model}\\\textbf{(randomly set)}\end{tabular}}           & 0                                  & 0.33            & 0.33              & 0.33                     \\ 
\cline{2-5}
                                                                                                                 & 0.95                               & 0.45            & 0.09              & 0.09                     \\ 
\hline
\multirow{2}{*}{\begin{tabular}[c]{@{}c@{}}\textbf{Naive Model}\\\textbf{(expert set)}\end{tabular}}             & 0                                  & 0.33            & 0.33              & 0.33                     \\ 
\cline{2-5}
                                                                                                                 & 0.95                               & 0.46            & 0.08              & 0.08                     \\ 
\hline
\multirow{2}{*}{\begin{tabular}[c]{@{}c@{}}\textbf{Updated Model with}\\\textbf{ Tuned Parameters}\end{tabular}} & 0                                  & 0.33                & 0.33                 & 0.33                        \\ 
\cline{2-5}
                                                                                                                 & 0.95                               & 0.5               & 0.1                 & 0.05                        \\
\hline
\end{tabular}
\end{table}

\paragraph{Case Study 1}

The first case focuses on the first honey-patch mitigation deployed in the game where both the attacker and the defender can take any possible actions, i.e., the attacker may choose to exploit any vulnerability or none at all, and the defender may choose to honey-patch any one or more vulnerabilities or choose to deploy no mitigation at all. As shown in Figure \ref{fig:attack_graph}, we are dealing with the case when Action Set 1 is available.

For the minimum weightage given to future actions and their corresponding expected payoffs using $\gamma=0$, each of the three models provide an equal probability to honey-patch each of the three vulnerabilities with the respective probabilities being 0.33 each. This is expected as independent of the transition probabilities, all three models share the same utility gains for honey-patching all three vulnerabilities, and the Q-value or the expected return does not depend on the transition matrix for $\gamma=0$, as shown in Equation \ref{eq:q_value}.

For the maximum weightage given to future actions and their corresponding expected payoffs using $\gamma=0.95$,
\begin{itemize}
    \item The randomly set Naive model provides a probability of 0.45 to honey-patch \textit{backup}, and an equal probability of 0.09 to \textit{sampleak} and \textit{exploit-market}.
    \item The expert set Naive model follows a similar trend and provides 0.46 to honey-patch \textit{backup} and 0.08 to the other two.
    \item The Updated Model with Tuned Parameters provides the highest probability of 0.5 to honey-patch \textit{backup}, a much smaller probability of 0.1 to \textit{sampleak}, and 0.05 to \textit{exploit-market}.
\end{itemize}

\textcolor{black}{Figures \ref{fig:uniform_model_s2} and \ref{fig:non_uniform_model_s2} show that the expected payoffs (Naive Model setups) are marginally higher than the actual payoffs received when the true knowledge about the attacker was used by the Updated Model with Tuned Parameters. Hence, we conclude that the defender's payoffs may not be as high as we may estimate them to be when modeling the attacker's behavior from the defender's perspective.}

\paragraph{Case Study 2}

The second case is when the attacker has captured either the real-flag or the honeypot flag for \textit{backup}, and the defender can choose to honey-patch either the second or the third vulnerability or choose to deploy no mitigation at all. Thus, this scenario represents the case when Action Set 2 is available.

For the minimum and the maximum weightage given to future actions and their corresponding expected payoffs using $\gamma=0$ and $\gamma=0.95$, respectively, all the three models give a 0.5 probability on deploying a honey-patch for \textit{sampleak} and 0 probability for deploying one on \textit{exploit-market}. \textcolor{black}{This can be seen in Figures \ref{fig:uniform_model_s5} and \ref{fig:non_uniform_model_s5} when the payoffs returned by all three models are equal.}

\paragraph{Case Study 3}

For each of the later states when the defender has the option either to honey-patch \textit{exploit-market} or use no mitigation at all, these states are cases when either the attacker captured 2 real-flags, 2 honeypot flags, or 1 of each kind for the first two vulnerabilities. Thus, this scenario represents the case when Action Set 3 is available.

For the minimum and maximum weightage given to future actions and their corresponding expected payoffs using $\gamma=0$ and $\gamma=0.95$, respectively, all the three models provide a 0.5 probability to honey-patch \textit{exploit-market}. \textcolor{black}{Figures \ref{fig:uniform_model_s8} and \ref{fig:non_uniform_model_s8} show equal payoffs for the defender in all three models.}
Note that in each of the case studies, the remaining probabilities (as the distribution over actions totals to 1.0) have been provided to the \textit{no mitigation} action for the defender.

% \begin{itemize}
%     \item The randomly set Naive model provides a 0.5 probability to recommend the defender to honey-patch \textit{exploit-market} and 0.5 to use no mitigation at all.
%     \item The expert set Naive model follows the exact same trend as the randomly set Naive model.
%     \item The Updated Model with Tuned Parameters follows the same trend as the model in (i).
% \end{itemize}

% For the maximum weightage given to future actions and their corresponding expected payoffs using $\gamma=0.95$,
% \begin{itemize}
%     \item The randomly set Naive model provides a 0.5 probability to recommend the defender to honey-patch \textit{exploit-market} and 0.5 to use no mitigation at all.
%     \item The expert set Naive model follows the exact same trend as the randomly set Naive model.
%     \item The Updated Model with Tuned Parameters follows the same trend as the model in (i).
% \end{itemize}

\subsection{Discussion}
\label{subsec:discussion}

When using uniform and non-uniform costs for the defender to deploy mitigation strategies, we note that the three models show slight variations only for the first three states of the game, particularly for the case when a high weightage is given to future gains or payoffs. Recall that these three states correspond to the following scenarios:

\begin{itemize}
    \item State $S_0$: when the attacker is at the starting state of the game and has acquired no flags so far.
    \item State $S_1$: when the attacker tried exploiting the first vulnerability and successfully obtained a real flag.
    \item State $S_2$: when the attacker tried exploiting the first vulnerability and was successfully trapped into the honeypot, thereby incurring a fake flag.
\end{itemize}

For all the other states, i.e., from state $S_3$ to state $S_9$, we note that all three models give the exact same utility returns. Also, irrespective of the state the defender is at the start when attempting the first vulnerability exploit, 
the three models provide a similar probability of $\approx$ 0.5 to deploy a honey-patch to mitigate the next possible exploit. Hence, we hypothesize
%believe that this can be attributed to the fact 
that \emph{the earlier the adversary is trapped in a honeypot, the better it is for the defender}.

The difference in the utility returns is the most prominent for state $S_2$, as shown in Figures \ref{fig:uniform_model_s2} and \ref{fig:non_uniform_model_s2}, where the attacker was trapped in the first honeypot of the game while exploiting the first vulnerability. The initial model with transition probabilities set by the expert dominates the other initializations and gives the highest return, followed by the initial model with probabilities set randomly, and the least returns are obtained through the model initialized using data. The most important observation here is that \emph{model parameters set randomly or by expert may not imitate the true model representative of the real-world attack scenario}. Hence, in the most important stages of the game, when the model results differ, defensive strategies tend to overestimate the returns from the randomly set or expert set model initializations as compared to the model we obtained from the real-world user studies.

\subsubsection{Evaluating the Hypotheses}
\label{subsubsec:hypotheses_evaluation}

In the first hypothesis, we assumed that the attacker, once trapped in a honeypot, may continue with the existing strategy without worrying about future honeypots. The results from the user studies show that none of the attackers received the observation, until informed explicitly about the honeypot flag, that they failed to get the real flag, thus verifying our expectation behind this hypothesis. From the equilibria comparison shown in Figures~\ref{fig:uniform_model} and \ref{fig:non_uniform_model}, we note that the payoffs for the defender are primarily equal to the payoffs obtained by the Naive Models, both with randomly set and expert set transition probabilities, thereby confirming that the adversary's behaviors did not deviate much from the system defender's expectations in this case.

In the second hypothesis, we assumed that the attacker, once trapped in a honeypot and knowing about the current state, will change the existing strategy to get out of the honeypot and exploit the future vulnerabilities with caution. Only in one instance, an attacker is able to escape the honeypot for \textit{sampleak} after the user study ends and the attacker is informed about the honeypot flag. In this case, factors such as attacker’s experience and expertise level, ease of exploiting the vulnerability and the effectiveness of the deployed honey-patch (e.g., an incomplete patch), all play an influential role. Due to these reasons, it is not easy to validate this hypothesis with 100\% confidence, and hence, we see this as one open direction for future research to investigate in detail.

\section{Related Work}
\label{sec:relatedwork}

Learning attack behaviors for a cybersecurity system has been a problem of relevant interest, particularly when designing decision making model frameworks for the defender. In an Internet of Things setting, Galinkin et al.~\cite{galinkin2021evaluating} classify attackers as risk-averse and risk-seeking to understand the suite of scenarios preferred by such adversaries. Assessing the different modalities that influence an attacker's decision making in real-world scenarios has proven to be a challenging task, and necessitates the requirement of better models that can capture such behavioral patterns more closely \cite{basak2018initial}. One of the recent attempts on learning attacker's behavior have been based on approximating the preferences and capabilities of the attacker based on previously collected data over network packets, to learn about the preferences, choices or capabilities of a potential adversary \cite{bounded_rationality_model}. However, understanding and collecting such data for adversaries, particularly when faced with decoy mitigation strategy, has not been analysed so far.

Do et al.~\cite{do2017game} survey existing game-theoretic techniques on cyber security and privacy challenges, and highlights the advantages and limitations from the design to implementation of defense systems. Such evaluations strongly encourage the need to utilize such effective modeling frameworks to fully comprehend the evolving security and privacy problems in cyberspace and to find viable solutions. On the other hand, cybersecurity exercises have also been popularly used as a platform to teach cyber security concepts, and also to conduct experiments to study, analyze and solve issues related to cybersecurity \cite{stransky2017lessons,schwab2019cybersecurity,salem2011design,salah2015teaching,mirkovic2012teaching,mases2019mixed,kavak2016characterization,aljohani2021conducting,sommestad2012cyber}.

The study setup presented in \cite{sommestad2012cyber} explores the use of cyber security exercises and competitions to obtain vital data on measuring the impact of mitigations against exploits and their corresponding success. In this work, our primary focus stayed on the deception-based mitigation of honey-patching vulnerabilities and how the adversary interacted when faced with such a scenario.

\section{Conclusion \& Future Work}
\label{sec:conclusion}

Cybersecurity exercises enable the collection of data on the interactions between attackers and a system defender, to gain insightful knowledge about the attacker's \emph{state}, which can be used to further to improve the strategy adopted by a defender when faced by a potential threat. In this work, we take on the challenge of analyzing closely these interactions in deception-based experimental setups where the adversary is faced by three different types of decoy traps. We started with a baseline game-theoretic framework where we manually set the probability distribution over the attacker's strategy, and update this model with the results collected using the cybersecurity exercise carried over a real-world CTF platform. We observe that the interactions between the defender and the adversary in the initial stages of the game makes a more significant difference in the total expected utility gain for the defender, than in the later stages of the interaction. Moreover, models initialized randomly or using subject-expert knowledge may also lead to the problem of overestimation for the defender's payoffs in certain scenarios. Since we have a constrained control over the different modalities influencing the adversarial behavior noted in these studies, we believe that gaining further knowledge on the attack behavior when faced with deception-based mitigation strategies holds promise for improving the defender's decision-making model. 

\subsubsection{Acknowledgements} This work was supported in part by U.S. ACC-APG / DARPA award W912CG-19-C-0003 and the U.S. Army Research Laboratory under Cooperative Agreement Number W911NF-13-2-0045. Any opinions, recommendations, or conclusions expressed are those of the authors and should not be interpreted as representing the official views or policies of the Department of Defense or the U.S. Government. Approved for Public Release, Distribution Unlimited. We would also like to thank Sailik Sengupta for his useful insights, helpful discussions and feedback on this work.

\bibliographystyle{unsrt}  
\bibliography{main}  

\end{document}